\documentclass[prb,twocolumn,superscriptaddress]{revtex4}
\usepackage{graphicx}

\begin{document}

\title{High intrinsic $ZT$ in InP$_3$ monolayer at room temperature}

\author{Shenghui Zhang}
\affiliation{Department of Physics, Shanghai Normal University, 100 Guilin Road, Shanghai 200232, P.R. China}
\affiliation{School of Materials and Energy, University of Electronic Science and Technology of China, Chengdu 610054, P.R. China}
\author{Xiaobin Niu}
\affiliation{School of Materials and Energy, University of Electronic Science and Technology of China, Chengdu 610054, P.R. China}
\author{Yiqun Xie}
\email{yqxie@shnu.edu.cn}
\affiliation{Department of Physics, Shanghai Normal University, 100 Guilin Road, Shanghai 200232, P.R. China}
\author{Kui Gong}
\affiliation{Hongzhiwei Technology (Shanghai) CO.LTD., 1888 Xinjinqiao Road, Pudong, Shanghai 201206, P.R. China}
\author{Hezhu Shao}
\email{hzshao@nimte.ac.cn}
\affiliation{Ningbo Institute of Materials Technology and Engineering, Chinese Academy of Sciences, Ningbo 315201, P.R. China}
\author{Yibin Hu}
\email{ybhu@mail.sitp.ac.cn}
\affiliation{State Key Laboratory of Infrared Physics, Shanghai Institute of Technical Physics, Chinese Academy of Sciences, Shanghai 200083, P.R. China}
\author{Yin Wang}
\affiliation{Department of Physics and International Centre for Quantum and Molecular Structures, Shanghai University, 99 Shangda Road, Shanghai 200444, P.R. China}

\begin{abstract}
Two-dimensional thermoelectric materials with a figure of merit $ZT$, which is greater than 2.0 at room temperature, would be highly desirable in energy conversion since the efficiency is competitive to conventional energy conversion techniques. Here, we propose that the indium triphosphide (InP$_3$) monolayer offers an extraordinary $ZT$ of 2.2 at 300 K by using quantum calculations within the ballistic thermal transport region. A remarkably low and isotropic phononic thermal conductivity is founded, which is due to flat lattice vibration modes. This low thermal conductivity takes a major responsibility to the impressively high $ZT$. Moreover, a large $ZT$ that is greater than 1.5 can be maintained,  even if a 1\% mechanic extension is applied on the lattice. These results suggest that the InP$_3$ monolayer is a promising candidate for low dimensional thermoelectric applications.
\end{abstract}

\maketitle

\section{Introduction}

Thermoelectric (TE) materials can realise direct energy conversion between heat and electricity, and have many potential applications in power generation and heat pumping\cite{Rowe1995}. The efficiency of thermoelectric materials is determined by the dimensionless figure of merit, $ZT=\frac{\sigma S^2 T}{\kappa_e+\kappa_p},$ where $\sigma$ is the electric conductivity, $S$ is the seebeck coefficient, $T$ is the absolute temperature and $\kappa_{p(e)}$ is the phononic (electronic) thermal conductivity\cite{PRL2009}. Great efforts have been put into finding thermoelectric materials with high $ZT$ value. So far, a maximum $ZT$ of 2.6 is reported in single crystal SnSe\cite{Nat2014}, and the $ZT$ value increases to 2.8 in n-tpye SnSe crystal\cite{Sci2018}.
Whereas at the nanoscale, quantum confinement can substantially reduce the phononic thermal conductivity, thus lead to a considerable enhancement of the thermoelectric efficiency in nanostructures. There are great interests in thermoelectric properties of the low-dimensional materials since the ground-breaking experiments which indicate that rough silicon nanowire can be an efficient thermoelectric material\cite{Nature451}.

Currently, there are intense focus on the thermoelectricity of the 2D materials, such as the graphenene, silicenene, black phosphorus and transition-metal dichalcogenides, due to their attractive electronic properties and thermal transport properties\cite{JAP2013, JCP2014, BP2016,Si2013,Si2014,2D_JPCC2016,MoS2017}. For example, a theoretical calculation has shown that the $ZT$ of silicene nanoribbon is close to 2.5 at 90 K.\cite{Si2013} A large $ZT$ of 2.8 at 800 K has been predicted for the monolayer SnSe. \cite{SnSe2017} Such excellent thermoelectric performances of 2D materials, however, were obtained when the temperatures were far away from the room temperature.

In reality, 2D thermoelectric materials with a high $ZT$ at room temperature will largely facilitate their utilizations, such as cooling and electricity generation. Unfortunately, the thermoelectric efficiency in the pristine 2D materials at room temperature is relatively poor, as the $ZT$ is typically less than 2.0. For instance, the $ZT$ of graphene is close to 0.01 \cite{graphene2013}, and those of the 2D black phoshporene and blue phosphorene are about 0.2 and 1.0, respectively.\cite{BP2016}  Moreover, 2D MX$_2$ ( M = Mo, W; X = S, Se) monolayers have $ZT<2$ at 300 K as predicated by theoretical calculation.\cite{JCP2014} So far, for the pristine 2D materials, the highest $ZT$ at room temperature is 2.15, which is calculated for the monolayer buckled antimonene\cite{JPCC2017}.

There are several methods that can be used to improve the thermoelectric efficiency in 2D materials, including defects and strain engineering, chemical doping and heterostucure\cite{JMCC2017}. For instance, the $ZT$ of graphene can be improved impressively to around 3.2 at room temperature using a layered structure with the phonon blocking materials\cite{GN2D_2018}. Besides, the $ZT$ of the antimonene can be largely improved from less than 0.1 to about 0.6 at room temperature by $n$-type doping \cite{Arse2017}. Even though, new 2D materials with a better thermoelectric efficiency at room temperature are still highly desirable in order to facilitate their real applications.

In this work, we investigated the thermoelectric properties of the InP$_3$ monolayer, a newly predicted 2D material with remarkable electronic properties\cite{InP2017a}, by using density functional theory combined with the non-equilibrium Green's functional formalisms (NEGF-DFT)\cite{Taylor}. We obtained a large $ZT$ of 2.2 at 300 K, due to the low thermal conductivity.

\section{Model and Methods}\label{sec:2}

\begin{figure}[htbp]
\centerline{\includegraphics[width=8cm]{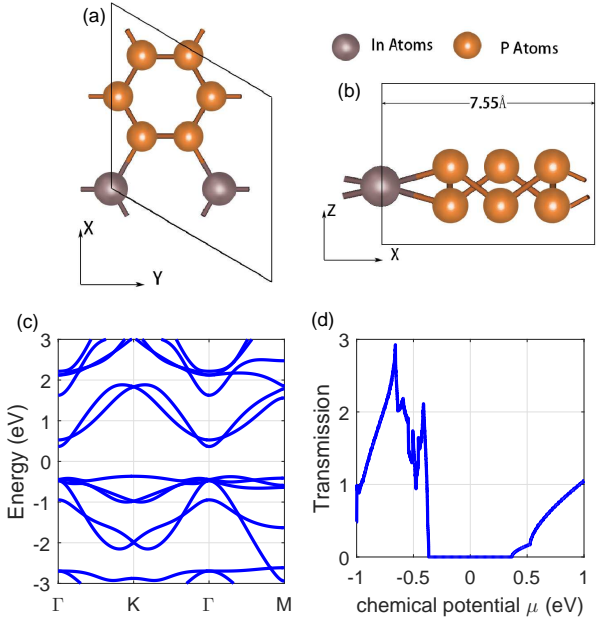}}
\caption{Top (a) and side (b) views of the primitive cell of the InP$_3$ monolayer. Gray and yellow spheres denote the In and P atoms, respectively. The electronic bandstructure (c) and electronic transmission spectrum (d) of the InP$_3$ monolayer.}
\label{fig1}
\end{figure}

The primitive cell of the InP$_3$ monolayer is presented in Figs.~\ref{fig1}(a,b), which is composed of two In atoms and six P atoms bonded via covalent interactions with a hexagonal structure, similar to the silicene. The lattice constant of the InP$_3$  monolayer is 7.55\AA\ optimized by VASP code\cite{vasp}, which is consistent with the previous theoretical calculation\cite{InP2017a}. Using this primitive cell, we calculated both the electronic and thermal transport properties of the InP$_3$ monolayer, and then obtained its thermoelectric properties. The details of the simulation methods are described in the following.

We optimize the lattice constant and obtain a force-constant calculated by VASP code\cite{vasp}, and construct the dynamic matrix for the phonon spectrum calculation. During the optimization, the atoms are fully relaxed until the maximum force is less than 0.005 eV/\AA. The plane wave was used for wave function expansion with a cutoff energy of 400 eV. The projector augmented-wave method \cite{PAW} was used for describing core electron. The PW91 version of the generalized gradient approximation (GGA) was used for the electron exchange and correlation functional\cite{PW_GGA}.
In structure relaxing, a $9 \times 9 \times 1$ k sampling was used. For calculating force-constant, a $3\times3\times1$ supercell was used, and k sampling reduced to $3\times3\times1$. Both the electronic and thermal transport properties were carried out using Nanodcal code\cite{nanodcal} within a NEGF-DFT theoretical method. In the calculation, $150\times 150 \times 1 $ k points were adopted. A double-zeta polarized (DZP) atomic orbital basis was used to expand all physical quantities, the exchanges and correlation was treated at the level of a generalized gradient approximation, and the atomic cores were defined using the standard norm-conserving nonlocal pseudopotentials. These calculation details were verified to provide accurate results.

Within the linear response limit, the electrical current and electrical thermal current can be defined by
\begin{equation}
I=\frac{2e}{h}\int T_e(E)(f_L(E)-f_R(E))dE
\end{equation}
\begin{equation}
I_Q=\frac{2}{h}\int T_e(E)(f_L(E)-f_R(E))(E-\mu)dE
\end{equation}
Here $T_e(\varepsilon)$ is the electronic transmission function, which can be calculated by the standard nonequilibrium Green's function method,
and $f(\varepsilon, \mu)=1/\{exp[(\varepsilon-\mu)/k_BT]+1\}$ is the Fermi-Dirac distribution function at the chemical potential $\mu$.

For ballistic electronic transport, electronic transmission function can be calculated as
\begin{equation}
T_e(E)=\mathrm{Tr}(G_e^r \Gamma_L G_e^a \Gamma_R)
\end{equation}
\begin{equation}
\Gamma_L = i(\Sigma_L^r-\Sigma_L^a),\Gamma_R = i(\Sigma_R^r-\Sigma_R^a)
\end{equation}
\begin{equation}
G_e^r = [ES - H - \Sigma_L^r-\Sigma_R^r]^{-1}
\end{equation}
Here, $G_e^r$ is retarded Green's function, $H$ and $S$ is hamiltonian and overlap matrix, $\Sigma_L^r$ and $\Sigma_R^r$ are self-energy from left and right semi-infinite leads.

From above equations, we can obtain electronic conductivity $\sigma$, Seebeck coefficient $S$ and thermal conductivity $\kappa_e$\cite{PRL2009,TM2006}
\begin{equation}
\sigma=e^2L_0/l
\end{equation}
\begin{equation}\label{eqS}
S=-\frac{L_1}{eTL_0}
\end{equation}
\begin{equation}
\kappa_e(T)=\frac{1}{Tl}\left(L_2-\frac{L_1^2}{L_0}\right)
\end{equation}
where $l$ is the device length, and $L_m(\mu)$ is given by
\begin{equation}
L_m(\mu)=\frac{2}{h}\int^{\infty}_{-\infty}d\varepsilon T_e(\varepsilon)(\varepsilon-\mu)^m\left(-\frac{\partial f(\varepsilon,\mu)}{\partial\varepsilon}\right)
\end{equation}

For ballistic phononic transport, the calculation of phonon transmission is similar to electron transmission. The difference is that electron transmission is calculated from hamiltonian matrix $H$ but phonon transmission is calculated from dynamic matrix $D$.
\begin{equation}
T_p(\omega)=\mathrm{Tr}(G_p^r \Gamma_L G_p^a \Gamma_R)
\end{equation}
\begin{equation}
\Gamma_L = i(\Sigma_L^r-\Sigma_L^a),\Gamma_R = i(\Sigma_R^r-\Sigma_R^a)
\end{equation}
\begin{equation}
G_p^r = [\omega^2 - D - \Sigma_L^r-\Sigma_R^r]^{-1}
\end{equation}

From above equations, the phonon transmission function $T_{p}(\omega)$ at frequency $\omega$ is calculated. And the phonon thermal conductivity can be written as\cite{PRL2009,TM2006}
\begin{equation}
\kappa_p(T)=\frac{\hbar^2}{2\pi K_BT^2l}\int^{\infty}_{0}d\omega \omega^2 T_p(\omega)\frac{e^{\hbar\omega/k_BT}}{(e^{\hbar\omega/k_BT}-1)^2}
\end{equation}

\section{Results and Discussion}

To obtain the thermoelectric properties of the InP$_3$ mononlayer, we first study its electronic transport properties. The electronic band structure is given in Fig.~\ref{fig1}(c), the top of valance bands ($E_{\mathrm{HOMO}}$) is -0.365 eV, the bottom of conduction bands ($E_{\mathrm{LUMO}}$) is 0.365 eV.
The work of Miao {\it et al} has pointed out that the top valence bands are mainly contributed by the P's 3$p$ and In's 5$p$ orbits. They have also related such dispersion of valance band with the Mexican-hat-like shape to the excellent magnetic, electrical and optical properties in the InP$_3$ mononlayer\cite{InP2017a}. Here, we focus on the thermoelectric property.
The characteristic of band structure, especially for the valance bands close to the Fermi energy, benefits largely to the high power factor of the InP$_3$ mononlayer. There are several bands including both heavy-hole and light-hole bands located at the $\Gamma$ point near Fermi energy. The light bands contribute to the high mobility. And the heavy bands indicate the large effective masses, which leads to a high Seebeck coefficient since the $S$ is proportional to the effective mass. Additionally, there are several heavy-hole bands, with similar energy value as those at $\Gamma$, at M and K points, which exhibits converged characteristic and multi-valley transport in material.
On the other hand, the top valance bands close to the Fermi energy ($\mu=0$) are very flat and intensive, as compared to the conduction bands. This corresponds to a high density of states and will lead to a large transmission coefficient for such valance bands.

Fig.~\ref{fig1}(d) gives the electronic transmission spectrum along the zigzag direction. A bandgap can also be observed, and moreover  the transmission coefficient for the valance bands (below the Fermi energy $\mu<0$) is indeed evidently larger than that for the conduction bands ($\mu>0$), which is consistent with the feature of the band structure.

\begin{figure}[htbp]
\centerline{\includegraphics[width=8cm]{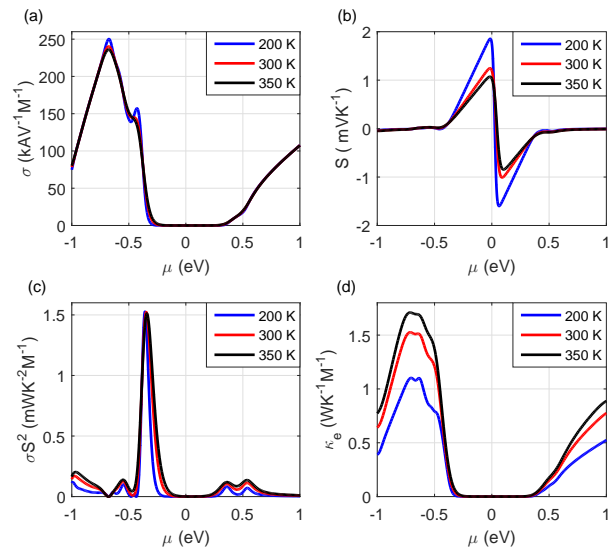}}
\caption{ (a) Electronic conductivity  $\sigma$, (b) Seebeck coefficient $S$, (c) power factor $\sigma S^2$, and (d) the thermal conductivity  of electrons $\kappa_e$ as a function of chemical potential $\mu$, respectively. }
\label{fig2}
\end{figure}

The electronic conductivity  $\sigma$ vanishes for chemical potential in the band gap at zero temperature. As the temperature is sufficiently high, transport is mediated by activated electrons and/or holes. Therefore, the nonzero conductivity  appears inside the bandgap due to the finite temperature. The electronic conductivity  $\sigma$ as a function of the chemical potential $\mu$ is shown in Fig.~\ref{fig2}(a) at 200 K, 300 K and 350 K, respectively. It shows that the $\sigma$ below the Fermi energy is evidently larger that above the Fermi energy, and shows a few peaks which correspond to those of the transmission spectrum (Fig.~\ref{fig1}(d)).
Figure \ref{fig2}(b) gives the Seebeck effect $S$ as a function of the chemical potential at temperatures of 200 K, 300 K, and 350 K, respectively. It can be seen that there are two peaks located below and above the Fermi energy, respectively, and inside the bandgap. Above (below) the Fermi energy, the $S$ is negative (positive) and its absolute value increases sharply and monotonously until a maxima is reached. For a positive (negative) chemical potential $\mu>0$ ($\mu<0$), the major charge carriers are electrons (holes), which flow from the left to the right at a  positive temperature difference ($\Delta T$) between the left and right. Therefore, the thermocurrent flows from the right to the left, and thus a positive (negative) voltage is needed to block the current, and hence the $S$ is negative (positive) according to its definition. In addition, the maxima of $S$ increases with the decreasing temperature, showing a sensitive dependence on the temperature. The appearance of the maxima of the $S$ can be ascribed to the relationship of the $S \propto L_1/L_0$, according to Eq.\ref{eqS}. Due to the factor $E-\mu$ in L$_1$, this function decreases with decreasing temperature faster than L$_0$ does. Therefore, the interplay of the $T$ dependence on the $L_0$ and $L_1$ functions leads to the maxima of the $S$.

The power factor $\sigma S^2$ at these temperatures is shown in Fig.~\ref{fig2}(c). A sharp peak of power factor is located at $\mu-E_{\mathrm{HOMO}} = 0.018$ eV when temperature is 300 K. Fig.~\ref{fig2}(d) gives the thermal conductivity  of electrons $\kappa_e$ as a function of chemical potential $\mu$, which increases with the increasing temperature.
\begin{figure}[htbp]
\centerline{\includegraphics[width=8cm]{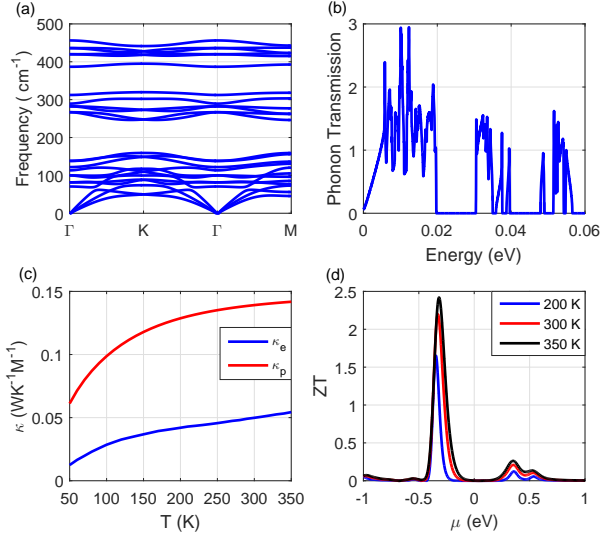}}
\caption{(a) Phonon spectrum of the InP$_3$ monolayer, and (b) the corresponding phonon transmission spectrum. (c) The phononic thermal conductivity  $\kappa_p$, and electronic thermal conductivity  $\kappa_e$. (d) The figure of merit $ZT$ as a function of chemical energy.}
\label{fig3}
\end{figure}

Having known the electronic transport properties of the InP$_3$ monolayer, we now investigate its thermal transport properties.
The phonon dispersion curves are shown in Fig.~\ref{fig3}(a), which is symmetrical with respect to the $\Gamma$ point for the $\Gamma\rightarrow$ K and $\Gamma \rightarrow$ M directions, suggesting an approximately isotropic thermal properties between the zigzag and armchair directions. There are twenty four curves in the phonon band structure that are contributed by the vibration modes of the two In atoms and six P atoms in the primitive  cell. The three lowest curves correspond to the three acoustic branches, that is, the z-direction acoustic (ZA) mode, in-plane transverse acoustic (TA) mode, and longitudinal acoustic (LA) mode. These three types of acoustic phonon modes have the highest group velocities among all phonon modes, thus they contribute most importantly to the thermal conductivity . The group velocity can be calculated by $\partial \omega /\partial k$, which are 1.04 kms$^{-1}$, 0.72  kms$^{-1}$ and 0.56  kms$^{-1}$ for the LA, TA, and ZA phonon modes, respectively. These group velocities are about one order lower than those of the monolayer black phosphorous \cite{BP2015}.   On the other hand, the whole dispersions of phonons are very flat and exhibit highly localized properties as shown in Fig.~\ref{fig3}(a), which demonstrates the low group velocities for most band branches. Note that a low phonon group velocity will lead to weak thermal transport capability, which means a lower thermal conductivity ,  and thus a higher $ZT$ is to be expected.

\begin{figure}[htbp]
\centerline{\includegraphics[width=8cm]{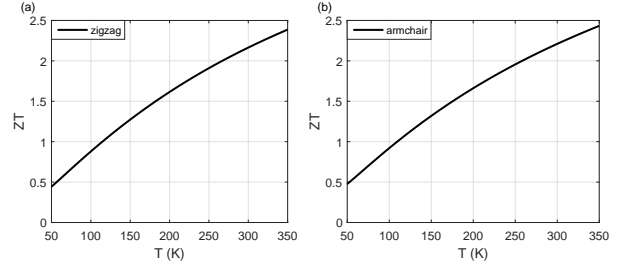}}
\caption{The variation of the figure of merit $ZT$ with temperature for the zigzag and armchair directions, respectively.}
\label{fig4}
\end{figure}

The phonon transport spectrum is shown in  Fig.~\ref{fig3}(b). It shows that the thermal transmission is larger for the the modes with energy lower than 0.02 eV. Importantly, these low energy phonon modes will give a major contribution to the phonon thermal conductivity  $\kappa_p$ for the temperature below 300 K. The dependence of the $\kappa_p$ on the temperature is shown in Fig.~\ref{fig3}(c). The $\kappa_p$ shows a monotonous increase with increasing temperature. In comparison, the thermal conductivity  contributed from electrons ($\kappa_e$) is also shown in the figure. The $\kappa_{p}$ is approximately 2.8 times larger than the $\kappa_e$ at 300 K, indicating that the phonon thermal conductivity  has a lager influence on the thermoelectric properties. Figure~\ref{fig3}(d) gives the figure of merit $ZT$ at 200 K, 300 K and 350 K, as a function of chemical potential. We found that there are several $ZT$ peaks located around the Fermi energy, and the maximum $ZT$ is located at $\mu-E_{\mathrm{HOMO}}=0.037$ eV with a large $ZT= 2.2$ at 300 K. This should be attributed to the large peak of the power factor $\sigma S^2$ at this chemical energy. Moreover, the maximum $ZT$ increases with the increasing temperature as shown in Fig.~\ref{fig4}(a), with the peak location shifting closer to the Fermi energy. In addition, we found that InP$_3$ monolayer has an isotropic thermoelectricity properties, as it has an approximately same $ZT$ along both zigzag and armchair directions, as shown in Fig.~\ref{fig4}.

The large $ZT$ (2.2) of InP$_3$ monolayer at room temperature outperforms that of other 2D materials ever reported, and is rather competitive in commercial applications. We found the phononic thermal conductivity  of the InP$_3$ monolayer is 0.14 W(mK)$^{-1}$, which is relatively less than several other 2D materials at 300K, such as silicene, germanene and SnSe.\cite{2D, SnSe2015} For example, the thermal conductivity of the single-layered SnSe sheet is 2.57 W(mK)$^{-1}$.\cite{SnSe2015} Thus the relatively lower thermal conductivity of the InP$_3$ monolayer takes an important responsibility to its large $ZT$.

\begin{figure}[htbp]
\centerline{\includegraphics[width=8cm]{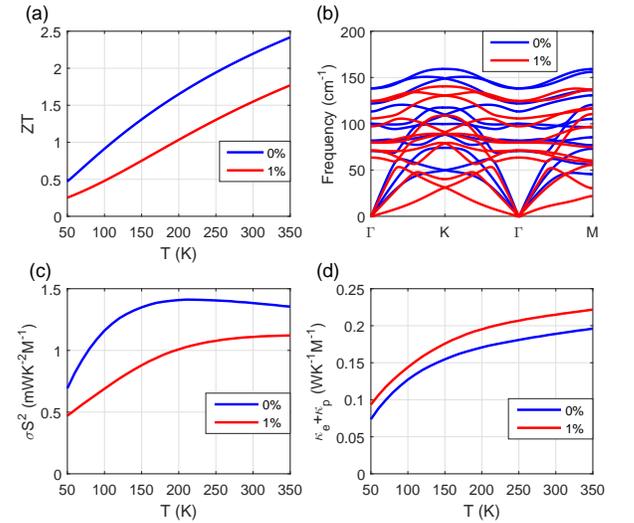}}
\caption{(a) the $ZT$, (b) the phonon bands, (c) the power factor, and (d) the thermal conductivity  ($\kappa_e+\kappa_p$) for the mechanic extension of 0\% and 1\%, respectively.}
\label{fig5}
\end{figure}

We now consider the influence of the mechanic strain on the figure of merit $ZT$. Under the  compression, the system is not stable and large imaginary frequency appears, while it is stable under the extension. In Fig.~\ref{fig5}(a), we show the $ZT$ of the InP$_3$ monolayer for the extension ratio of 0\% and 1\%. It shows clearly that the $ZT$ drop obviously with the extension, which is 1.55 at the 300 K for the 1\% mechanic extension strain. However, the $ZT$ under mechanic extension strain is still considerably large, as it is greater than 1.5. Fig.~\ref{fig5}(b) shows the phonon bands for 0\% and 1\%, respectively. From the results, we can find that the phonon frequency is decreased under the mechanic extension throughout. The decrease of the $ZT$ under the mechanical extension can be understood from the variation in the thermal conductivity and the power factor $\sigma S^2$. Fig.~\ref{fig5}(c) shows that under the 1\% mechanic extension strain, the power factor decreases significantly, as compared to the case without strain, which will decrease the $ZT$. On the other hand, thermal conductivity increases a little under strain (Fig.~\ref{fig5}(d)), which will also decrease the $ZT$. These two effects lead to decrease the $ZT$ finally, as the effect in the power factor is more severely than that of the thermal conductivity.

It should be noted that the transport of the thermal properties is calculated within the ballistic transport region. It means that other factors like phonon-phonon scattering effects and electron-phonon scattering effects which are important in higher temperatures are not considered. These factors can be safely eliminated below the Debye temperature. We estimated that the Debye temperature is about 380 K, which is obtained by fitting the Debye formula\cite{Shao2016epl},
\begin{equation} \label{equation:c2}
C_v=9Nk_B(\frac{T}{\Theta_D})^3\int_{0}^{\Theta_D/T}\frac{x^4 e^x dx}{(e^x-1)^2},
\end{equation}
where $x=\hbar\omega/k_B T$. The isometric heat capacity $C_v$ can be obtained by
\begin{equation} \label{equation:c1}
C_v=\sum_{n,\mathbf{q}} k_B(\frac{\hbar\, \omega_n(\mathbf{q})}{k_B T})^2 \frac{e^{\hbar\, \omega_n(\mathbf{q})/k_b T}}{(e^{\hbar\, \omega_n(\mathbf{q})/k_B T}-1)^2},
\end{equation}
where $\hbar$ is the reduced Planck constant, $k_B$ is Boltzmann constant, $T$ is the temperature, and $\omega_n(\mathbf{q})$ is the phonon frequency of the $n$th branch with wave vector $\mathbf{q}$. We could solve Eq.(\ref{equation:c2}) numerically, and get the temperature-dependent Debye temperature. We determine the Debye temperature, which is the value at the temperature point where the heat capacity is equal to the half of Dulong and Petit value.
Above the Debye temperature, all the phonon modes are activated, and the phonon-phonon scattering would dominate in determining the behavior of temperature-dependent thermal conductivity, which would decrease with temperature according to the $1/T$ relation. More importantly, due to the 2D property of present system of InP$_3$, the phonon-surface scattering in decreasing the heat conductivity outweighs the importance of phonon-phonon scattering, which is just like that in many film systems\cite{Tritt2003}. Then the anharmonic effects is of limited importance, and the ballistic transport is expected to be good in describing the behavior of heat conductivity  of the InP$_3$ monolayer at lower temperature. Therefore, the high $ZT$ of 2.2 at 300 K for the InP$_3$ monolayer is reliable in this work.

\section{Conclusions}
In summary, we have studied the intrinsic thermoelectrical properties of the InP$_3$ monolayer using the quantum transport calculations within the ballistic transport region. Our calculation shows that there are the converged valance band structures near the Fermi energy, which benefits the high power factor in the InP$_3$ monolayer. More importantly, the low group velocities of phonons leads to the low thermal conductivity. And it leads to high thermoelectric performance of the InP$_3$ monolayer. Moreover, the $ZT$ can be greater than 1.5, even if a certain mechanic extension ($\leq 1\%$) is applied on the lattice. Our work gives an insight to searching high $ZT$ materials at room-temperature.

\section{Acknowlegements}
This work was supported by the National Natural Science Foundation of China under Grants Nos. 11404348, 11504395 and 51871156.

\bibliography{reference}
\bibliographystyle{unsrt}

\end{document}